\documentclass[aps,prc,showpacs,twocolumn]{revtex4}
\usepackage{graphics}
\usepackage{dcolumn}
\usepackage{bm}
\begin{document}
\title{ Behaviour of the $\Lambda N$- and $\Lambda N N$- potential 
strengths in the $_\Lambda^5$He hypernucleus}
\author{A. A. Usmani}
\email{ anisul@iucaa.ernet.in}
\affiliation
{Department of Physics, Aligarh Muslim University, Aligarh 202 002, India}
\affiliation
{Inter University Centre for Astronomy and Astrophysics (IUCAA), 
Ganeshkhind, Pune-411 007, India}
\author{F. C. Khanna}
\email{ khanna@phys.ualberta.ca}
\affiliation
{Physics Department, Theoretical Physics Institute,
University of Alberta, Edmonton,  T6G 2J1, Canada}
\affiliation
{TRIUMF, 4004, Wesbrooke Mall, Vancouver, British Columbia, V6T 2A3, 
Canada}

\date{\today}
\begin{abstract}
Variational study of the $_\Lambda^5$He hypernucleus is presented using 
a realistic Hamiltonian and a fully correlated wave function including
$\Lambda N$ space-exchange correlations.  Behaviour of  $\Lambda$-separation 
energy ($B_\Lambda$) with two- and three- baryon
potential strengths is thoroughly investigated.  Solutions for
these potential strengths giving experimental $B_\Lambda$ are presented.
\end{abstract}

\pacs{21.80.+a, 21.10.Pc, 13.75.Ev, 13.75.Cs}
\keywords{Hypernuclei, Variational Monte Carlo, realistic interactions, 
$\Lambda N$ space exchange correlation}

\maketitle
Recently, a variational study~\cite{AAU06} of the $_\Lambda^5$He 
hypernucleus has been performed with a realistic Hamiltonian and a fully 
correlated wave function (WF). The WF takes into account all relavant dynamical 
correlations induced by the two- and three- baryon potentials and the 
$\Lambda N$ space-exchange correlation ($SEC$) that arises due to the 
space-exchange potential.  
The findings of the investigation suggest that no realistic study 
ignoring $SEC$ is fair as it significantly affects every physical observable 
like energy breakdown, $\Lambda$-separation energy ($B_\Lambda$), nuclear core 
polarization ($NCP$), point proton radius and density profiles. The effect is 
found more evident in the $_{\Lambda\Lambda}^{\:\:\:6}$He double-$\Lambda$ 
hypernucleus~\cite{AAU206}.  The ground-state energy of the hypernucleus 
($E$) or the $\Lambda$-separation energy ($B_\Lambda=E_{^4{\rm He}}-E$) 
depends on the strengths of the potentials involved in the Hamiltonian. 

A realistic Hamiltonian $H$ of the hypernucleus, is written as 
a sum of the Hamiltonians due to the nuclear core ($NC$)  of the hypernucleus 
($H_{NC}$) and due to the $\Lambda$-baryon ($H_\Lambda$), 
\begin{eqnarray}
H&=&H_{NC}+H_\Lambda, \\
H_{NC}&=&T_{NC} +\sum_{i< j} v_{ij} +\sum_{i< j < k} V_{ijk}, \\
H_\Lambda &=&T_\Lambda +\sum_i v_{\Lambda i}+\sum_{i< j} V_{\Lambda ij}.
\end{eqnarray}
Here, subscripts $i,j$ and $k$ refer to nucleons.
For the $S=0$ sector,  we use Argonne $v_{18}$ $NN$ potential~\cite{AV18} and 
Urbana type $NNN$ potential~\cite{NNN-IX,NNN}, which successfully explain
the nuclear energy spectra and are well established. 
However, for the  $S=-1$ sector, $\Lambda N$ and $\Lambda NN$ potential 
strengths are yet to be determined. The dependence of energy on these
strengths is the theme of this work. 

Variation in any of the potential strengths would directly affect the 
expectation value of the respective potential.  It would also affect the WF 
as correlations are a  solutions of these potentials. Moreover, there are 
sensitivities to operators among various terms of the Hamiltonian and the 
correlation functions. Correlations like $SEC$ bring-in changes in the 
density profiles which affect even the central pieces of the energy 
breakdown. The basic ingredients like these strengths, therefore affect the 
ground-state energy collectively.  It would not be possible to perform a 
proper study for a particular potential strength ignoring others. Thus, in order 
to pin down these potential strengths, we have to handle them all together. 
The energy of $_\Lambda^5$He is thoroughly investigated along  these lines.
Such a study of all the s-shell single- and double-$\Lambda$ hypernuclei
may help to pin down these strengths, which, in turn, may resolve the outstanding 
$A=5$ anomaly~\cite{DHT72,Gal75,Hungerford84} with no  additional effort.
Thus the present investigation is a step forward to answer, 
(i) whether we can successfully 
reproduce the hypernuclear energy spectra using these potentials without invoking 
the underlying Quantum Chromodynamics?  (ii) and whether there is a possibility 
of physical existence of a bound ($I=0,J=1^+$) $_{\Lambda\Lambda}^{\;\;\;4}$H 
hypernucleus which has been recently conjectured~\cite{Ahn01}?  This would  be 
helpful in studying the heavy hypernuclei specially $_{\:\:\:\: \Lambda}^{209}$Pb 
whose core density matches the nuclear matter density. Hence, this would lead
us to investigate, in detail, the physics of charmed and bottom 
hypernuclei~\cite{Tyapkin,Dover,Khanna} as well as to the nature and structure 
of neutron stars. 

\begin{figure}
\includegraphics{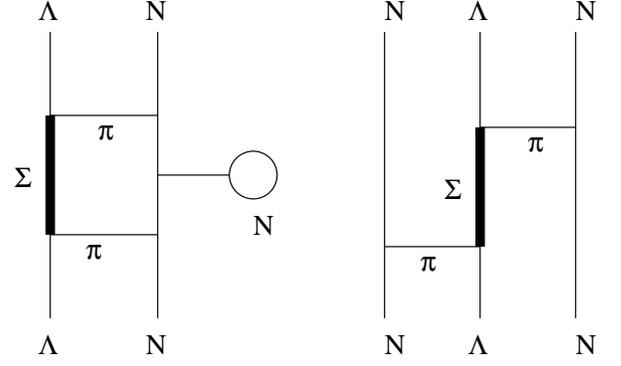} 
\caption{\label{fig1} 
Diagarm representing $V_{\Lambda NN}^D$ and $V_{\Lambda NN}^{2\pi}$.
}
\end{figure}

The $\Lambda NN$ potential arises from projecting out $\Sigma$, $\Delta$, 
etc., degrees of freedom from a coupled channel formalism.
This is written as a sum of two terms,
$V_{\Lambda ij}=V_{\Lambda ij}^D + V_{\Lambda ij}^{2\pi}$,
as in Fig.~\ref{fig1}.
The dispersive potential $V_{\Lambda ij}^D$, arising from the  suppression
mechanism owing to $\Lambda N -\Sigma N$ coupling~\cite{BR70,BR71,Rozynek79,
UB99}, is written including the  explicit spin dependence as~\cite{BU88} 
\begin{equation}
\label{VLNND}
V_{\Lambda ij}^{D}=
W^{D}T_{\pi}^{2}(m_\pi r_{\Lambda i})T_{\pi}^{2}(m_\pi r_{\Lambda j})
  [1+\mbox{\boldmath$\sigma$}_{\Lambda}\cdot(
\mbox{\boldmath$\sigma$} _{i}+ \mbox{\boldmath$\sigma$}_{j})/6].
\end{equation}
However, spin term is too weak for spin zero core nucleus. Here, $W^D$ is the 
strength.  The $V_{\Lambda ij}^{2\pi}$ is a two-pion exchange  
attractive potential. Neglecting  higher partial waves it is written as a sum of 
two terms representing p- and s-wave $\pi-N$ scatterings, 
$V_{\Lambda ij}^{2\pi}=V_{\Lambda ij}^{P} +V_{\Lambda ij}^S$
as in Ref.~\cite{Bhaduri67}. 
The explicit form of these potentials are%
\begin{equation}
\label{VP}
V^{P}_{\Lambda ij}=-\left(C^{P}/6\right)
(\mbox{\boldmath$\tau$}_{i}\cdot
\mbox{\boldmath$\tau$}_{j})
\{X_{i\Lambda},X_{\Lambda j}\},
\end{equation}
and
\begin{eqnarray}
\label{VS}
V^{S}_{\Lambda ij}&=&C^{S}
Z(m_\pi  r_{i\Lambda})Z(m_\pi r_{j\Lambda })
\mbox{\boldmath$\sigma$}_i\cdot {\hat{\bf r}}_{i\Lambda }
\mbox{\boldmath$\sigma$}_j\cdot {\hat{\bf r}}_{j\Lambda}
\mbox{\boldmath$\tau$}_i\cdot\mbox{\boldmath$\tau$}_j \nonumber \\
&\equiv& C^S O_{\Lambda ij}^S
\end{eqnarray}
with
\begin{equation}
\label{XiL}
X_{\Lambda i}=( \mbox{\boldmath$\sigma$}_{\Lambda}\cdot  
\mbox{\boldmath$\sigma$}_{i} )Y_{\pi}(m_\pi r_{\Lambda i})+S_{\Lambda i}
T_{\pi}(m_\pi r_{\Lambda i})
\end{equation} 
and
\begin{equation}
Z(x)=\frac{x}{3} [Y_\pi(x)-T_\pi(x)].
\end{equation}
It may be  expressed as  generalised tensor-tau type 
operators $(\mbox{\boldmath$\sigma$}_i \cdot{\bf r}_{\Lambda i})
(\mbox{\boldmath$\sigma$}_j\cdot{\bf r}_{\Lambda j})$,
$(\mbox{\boldmath$\sigma$}_i \cdot{\bf r}_{\Lambda i})
(\mbox{\boldmath$\sigma$}_j\cdot{\bf r}_{\Lambda i})$,
$(\mbox{\boldmath$\sigma$}_i \cdot{\bf r}_{\Lambda j})
(\mbox{\boldmath$\sigma$}_j\cdot{\bf r}_{\Lambda j})$,  and
$(\mbox{\boldmath$\sigma$}_i\cdot \mbox{\boldmath$\sigma$}_j)$
followed by $(\mbox{\boldmath$\tau$}_i\cdot \mbox{\boldmath$\tau$}_j)$,
thus has a strong tensor dependence.  
In the above expressions, $S_{\Lambda i}$
is the tensor operator, $Y_\pi (x)$ is the  Yukawa function
\begin{equation}
\label{eq3}
Y_\pi(x)=\frac{e^{-x}}{x} \xi_Y(r) ,
\end{equation} 
and $T_\pi (x)$ is the one-pion exchange tensor potential
\begin{eqnarray}
\label{Tpi}
T_{\pi}(x)&=&\left(1+ \frac{3}{x}+ \frac{3}{x^2} \right)
\frac{e^{-x}}{x} \xi_T(r). 
\end{eqnarray}
Here,  $\xi_Y(r)$ and $\xi_T(r)$ are short-range cut-off functions, 
\begin{equation}
\xi_Y(r)=\xi_T^{1/2}(r)=\left(1-e^{-cr^2}\right).
\end{equation}
Here, $c=2.0$ fm$^{-2}$ is a cut-off parameter
and subscripts $i,j$ and $\Lambda$ refer to two nucleons 
and a $\Lambda$ in the triplet ($\Lambda ij$).
The  $C^{P}$ and  the $C^{S}$  are the strengths of 
$V^{P}_{\Lambda ij}$ and $V^{S}_{\Lambda ij}$, respectively.
%These strengths  are not fixed. 
%The $V^{S}_{\Lambda ij}$ potential is a very weak term compared to the 
%$V^{P}_{\Lambda ij}$ potential as it is strongly suppressed like $NNN$ s-wave 
%potential.  
The latter is a very weak term compared to the former.
Its strength is not known experimentally. However, 
we may make a qualitative theoretical estimate for it by comparing the
strengths of modern $NNN$ potentials obtained using $SU(3)$ symmetry. 
Using chiral perturbation theory, Friar {\it et al.}~\cite{Friar99} have 
compared the modern $NNN$ potentials, namely: (i) Tucson-Melbourn~\cite{Coon79},
(ii) Brazil~\cite{Coelho83}, (iii) Ruhr~\cite{Eden93} and 
(iv) Texas~\cite{Kolck94}
containing a $\sigma$-term for $\pi-N$ scattering. They also consider  
the Fujita-Miyazawa force~\cite{Fujita57} dropping s-wave pions 
and Urbana-Argonne model~\cite{NNN} with additional isospin- and spin-
independent components added  to the Fujita-Miyazawa force.

In the Tucson-Melbourne (TM) model, the s-wave NNN force is written in the form
\begin{equation}
B({\bf r}_{ij}, {\bf r}_{jk})\{
\mbox{\boldmath$\tau$}_i \cdot\mbox{\boldmath$\tau$}_j,
\mbox{\boldmath$\tau$}_j \cdot\mbox{\boldmath$\tau$}_k\}
\{(S_{ij}+\mbox{\boldmath$ \sigma$}_i\cdot\mbox{\boldmath$\sigma$}_j)
(S_{jk}+\mbox{\boldmath$ \sigma$}_j\cdot\mbox{\boldmath$\sigma$}_k) \},
\end{equation}   
where, $B({\bf r}_{ij}, {\bf r}_{jk})$ has several terms given in 
Ref.~\cite{NNN}.  Recently, this has been expressed retaining only the term 
with pion-exchange-range functions as~\cite{SCP01}
%\begin{eqnarray}
%\label{AS}
%&A^S=\left(\frac{f_{\pi NN}}{4\pi}\right)^2a^\prime m_\pi^2, \\
%&O^{S}_{ijk}= \sum_{cyc}
%Z(m_\pi  r_{ij})Z(m_\pi r_{jk })
%\mbox{\boldmath$\sigma$}_i\cdot {\hat{\bf r}}_{ij }
%\mbox{\boldmath$\sigma$}_k\cdot {\hat{\bf r}}_{kj}
%\mbox{\boldmath$\tau$}_i\cdot\mbox{\boldmath$\tau$}_k. \nonumber \\
%\end{eqnarray}
\begin{equation}
\label{AS}
A^S=\left(\frac{f_{\pi NN}}{4\pi}\right)^2a^\prime m_\pi^2, 
\end{equation}
\begin{equation}
O^{S}_{ijk}= \sum_{cyc}
Z(m_\pi  r_{ij})Z(m_\pi r_{jk })
\mbox{\boldmath$\sigma$}_i\cdot {\hat{\bf r}}_{ij }
\mbox{\boldmath$\sigma$}_k\cdot {\hat{\bf r}}_{kj}
\mbox{\boldmath$\tau$}_i\cdot\mbox{\boldmath$\tau$}_k. 
\end{equation}
The parameter $a^\prime$, whose value ranges from $-0.51/m_\pi$ to $-1.87/m_\pi$,
is listed in Ref.~\cite{Friar99}. The TM value $a^\prime=-1.03/m_\pi$ gives the 
strength $|A^{S}| \approx$0.8 MeV.  However, in many others it is assumed to 
have a value of 1.0 MeV.  

Comparing TM model with the Eq.~\ref{VS} for $\Lambda NN$ potential, 
one may write an identical structure for both s-wave $\Lambda NN$ 
and $NNN$ potentials as following
\begin{equation}
C^S O^S_{\Lambda jk}\equiv A^S O_{ijk}.
\end{equation}
This directly relates $C^{S}$ in the 
strange sector to $A^{S}$ in the non-strange sector. 
Since $\Lambda- N$ mass difference is small compared to the $\Delta- N$ 
mass difference, $\Lambda NN$ potential of $S=-1$ sector is stronger than its 
non-strange counterpart $NNN$ potential~\cite{Bhaduri67} of $S=0$ sector. 
%Its long range part has the same range as $\Lambda N$ force. 
This provides stronger strengths in the case of $\Lambda NN$ potential 
compared to the $NNN$ potential.  We, therefore, expect that the value of $C^{S}$ 
would be more than 1.0 MeV, and is taken to be 1.5 MeV. 	               

The charge symmetric $\Lambda N$ potential~\cite{BUC84,Lagaris81} reads as
\begin{equation}
\label{vLN}
v_{\Lambda N}(r)=v_0(r) (1-\varepsilon+\varepsilon P_{x})
+(v_{\sigma}/4)T_{\pi}^{2}(m_\pi r)
\mbox{\boldmath$\sigma$}_{\Lambda}\cdot\mbox{\boldmath$\sigma$}_{N}.
\end{equation}
The first term includes direct potential ($v_0(r)=v_c(r)-v_{2\pi}(r)$)
and  space-exchange potential ($\varepsilon v_0(r)(P_x-1)$). Here,
$\varepsilon$ determines the odd-state potential, which is the 
strength of the space-exchange potential relative to the direct potential.  
%Its value is poorly estimated from the $\Lambda p$ forward-backward asymmetry
%that ranges from 0.1 to 0.38~\cite{UB99}.
Its estimate from the $\Lambda p$ forward-backward asymmetry is poor that
ranges from 0.1 to 0.38~\cite{UB99}.
The potential $ v_c(r)=W_c/[1+{\rm exp}\{(r-R)/ar\}]$ is the Saxon-Woods repulsive 
potential, with $W_c=2137$ MeV, $R=0.5$ fm and $a=0.2$ fm, and 
$v_{2\pi}=\overline{v}T_{\pi}^{2}(m_\pi r)$ is the two-pion attractive potential. 
The constants, $\overline{v}=(v_{s}+3v_{t})/4$ and $v_{\sigma}=v_{s}-v_{t}$, are 
respectively the spin-average and spin-dependent strengths, with $v_{s(t)}$ 
the singlet(triplet) state potential depth. 

\begin{table} 
\caption{\label{tab0}
$\Lambda N$ potential strengths in units of MeV.}
\begin{ruledtabular}
\begin{tabular}{lcccc}
 & $v_s$ &  $v_t$  & $\overline{v}=(v_s+3v_t)/4$  &  $v_\sigma =v_s-v_t$  \\
\hline
$\overline{v}1$ &
  6.33 & 6.09  & 6.15   & 0.24  \\
$\overline{v}2$ &
  6.28 & 6.04  & 6.10   & 0.24  \\
$\overline{v}3$ &
  6.23 & 5.99  & 6.05   & 0.24  \\
\end{tabular}
\end{ruledtabular}
\end{table}

We perform variational Monte Carlo study to calculate the ground-state energy, 
$E={\langle \Psi|H|\Psi\rangle} /{\langle\Psi| \Psi\rangle}$, 
where $\Psi$  is the WF of the hypernucleus. The computational details are 
available in Ref.~\cite{AAU06}.  
For the spin-zero core nucleus, the expectation value of the spin part of the 
$\Lambda N$ potential is negligibly small~\cite{AAU06, AAU03,UPU95,AAU95}. So is 
the s-wave part of $\Lambda NN$ potential.  Moreover, correlations 
induced by them are too weak to offer any significant change in the energy.  
Thus, we choose a reasonable strength for these two. The energy is very sensitive 
to changes in the $\Lambda N$ potential strengths $\overline{v}$ and $\varepsilon$. 
They implicitly appear in the WF through the $\Lambda N$ central and the $SEC$ 
correlations. The $\Lambda NN$ potential and its correlations involving
$C^P$ and $W^D$ play an important role. Therefore, 
$E$ or $B_\Lambda$ are sensitive to the strengths  $\overline{v}$, 
$\varepsilon$, $C^P$ and $W^D$.

The value of $\overline {v}\approx6.15(5)$ MeV 
is found consistent with the low energy $\Lambda p$ scattering data~\cite{BU88}. 
We use three different sets of $v_s$ and $v_t$, which give three different
values of $\overline{v}$ and a constant $v_\sigma$ as in Table~\ref{tab0}, 
referred to as $\overline{v}1$, $\overline{v}2$ and $\overline{v}3$. 
For all these, we choose three values of $\varepsilon$
in the range from 0.1 to 0.38 as mentioned before. These are  
0.1, 0.2 and 0.3. 
Results for these values are given in Tables~\ref{tab1}, \ref{tab2} and 
\ref{tab3}.

The correlations induced by different components of $\Lambda NN$ potential is 
written using scaled pair distances ($\overline{r}$) and  a variational 
parameter $\delta^m$ as in Ref.~\cite{AAU03},
\begin{equation}
\label{Ulij}
U_{\Lambda ij}=\sum_m U_{\Lambda ij}^m= \sum_{m=D,P,S}\delta^m V^{m}
(\overline{r}_{\Lambda i},\overline{r}_{ij}, \overline{r}_{j\Lambda}). 
\end{equation}
The $\langle V^D_{\Lambda ij}\rangle$ obeys a  linear behaviour,
$\partial V^D_{\Lambda ij}/\partial W^D=\partial E/\partial W^D$=constant at a 
fixed $\varepsilon$ as it is not sensitive to the operators but only to $SEC$ and 
hence to $\varepsilon$. As is obvious from Eq.~\ref{Ulij}, a change in the strength 
$C^P$ offsets the WF. We observe that along with its own 
correlation parameter $\delta^P$  a couple of other parametres are found 
to change with $C^P$.  But the repulsive $\Lambda NN$ correlation, 
$U_{\Lambda NN}^{D}$, remains invariant. 
We perform calculations for a wide range of $C^P$ starting from 0.5 MeV and upto
2.0 MeV. Therefore, enhancements in the attraction due to increasing $C^P$ needs 
to be balanced by an appropriate increase in $W^D$. 
For every independent calculation, we tune the WF afresh and adjust the 
repulsive strength $W^D$ in order to reproduce the experimental $B_\Lambda$.  
For the new $W^D(new)$, we may easily obtain new $\delta^D(new)$ using
\begin{equation}
\delta^D (old) W^D(old)= \delta^D (new) W^D (new),
\end{equation}
as $U^D_{\Lambda NN}$ is constant with $C^P$.
Thus $\delta^D$ decreases in the same proportion $W^D$ increases.  The same is not 
true in the case of $\delta^P$, which is found to be alomst constant even if we 
multiply $C^P$ by a factor of 4.  This is because of the sensitivity of 
$V^P_{\Lambda ij}$ to its correlation, which is so strong that
the attraction, $\langle V_{\Lambda ij}^{2\pi}\rangle$, increases more than 
12 times for the corresponding 4 times increase in $C^P$.
This quadratic behaviour is observed for all the $\varepsilon$ and all the 
$\overline{v}$.  The respective increase in the $NC$ part of the energy 
($E_{NC}$) is  about 4 MeV. The $T_\Lambda$ and $v_{\Lambda i}$ also  exhibit 
considerable change due to the variation in $C^P$.  
\begin{figure}
\includegraphics {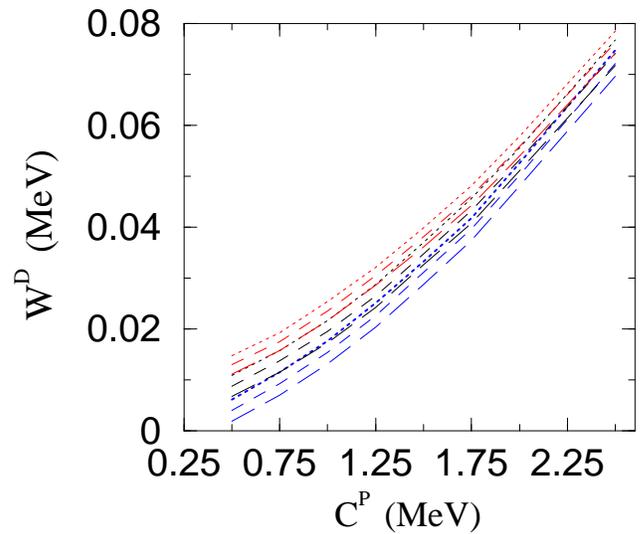}
\caption{\label{fig2} 
Curves show the set of strengths giving  $B_\Lambda^{exp}$.
The dotted, dashed and long dashed lines represent $\varepsilon$=0.1, 0.2 and 0.3, 
and the  red, black and blue  colors represent 
$\overline{v}1$, $\overline{v}2$ and $\overline{v}3$, respectively. 
} 
\end{figure}

Solutions of all these strengths reproducing $B_\Lambda^{exp}$ are plotted
in Fig.~\ref{fig2}. Thus, following the curves one may find potential 
strengths that reproduce $B_\Lambda^{exp}$. 
For every $\overline{v}$ and at a fix value of $C^P$, we observe a linear 
realtionship, ${\partial W^D}/{\partial \varepsilon}\approx c$, which is a 
consequence of two other linear relationships: 
(i)
$ {\partial E} / {\partial \varepsilon}= 
-{\partial B_\Lambda} / {\partial \varepsilon} \approx c_1$   
(ii) and
$ {\partial V^D_{\Lambda ij}} / {\partial W^D} =
- {\partial E} / {\partial W^D}= 
 {\partial B_\Lambda} / {\partial W^D} \approx c_2$.
The slope, $c=c_1/c_2$ is found to  increases with $C^P$,
but only slightly (Table~\ref{tab5}).  Curves representing 
different $\overline{v}$ are found to get closer at higher $C^P$.  Because, the 
dependence of energy on $C^P$ as well as on  $W^D$ varies with 
$\overline{v}$. To match an increase in the attractive $C^P$
we require a larger increase in  the repulsive $W^D$ for smaller $\overline{v}$.

This study has established the value of parameters appearing in the two- and 
three-body potentials in the strange sector. Furthermore, range of variations
of these parameters is now, atleast roughly, known. In order to establish 
these parameters over a range of hypernuclei, numerous light and heavy 
hypernuclei have to be studied. This would help us to understand the variation of
$\Lambda N$ and $\Lambda NN$ potentials as a function of density. Such a study
would be in parallel to that of $NN$ and $NNN$  potentials in nuclear systems
where density variation plays an important role. A proper understanding of
hypernuclei would help us to clarify the behaviour of the nuclear forces in
the strange sector. Ultimately, a detailed study may lead us to a clarification
of the role of QCD in determining the potential strengths. New results expected 
from Japan Hadron Facility would help to sort out these questions in the near 
future.

The work is supported under Grant No. SP/S2/K-32/99 awarded 
to AAU by the Department of Science and Technology, Government of India.
FCK thanks  NSERCC for financial support.
\begin{turnpage}
\begin{table}
\caption{\label{tab1} Energy breakdown for the set 
$\overline{v}1$ ($\overline{v}=6.15$ and $v_\sigma=0.24$).  
All quantities are in units of MeV except for $\varepsilon$.
}
\begin{ruledtabular}
\begin{tabular}{lcccccccccccc}
   &\multicolumn{4}{c}{ $\varepsilon=0.1$ }
   &\multicolumn{4}{c}{ $\varepsilon=0.2$ }
   &\multicolumn{4}{c}{ $\varepsilon=0.3$ } \\
 &   $C^p=.5$   &  $C^p=1.0$  &  $C^p=1.5$ & $C^p=2.0$  
 &   $C^p=.5$   &  $C^p=1.0$  &  $C^p=1.5$ & $C^p=2.0$  
 &   $C^p=.5$   &  $C^p=1.0$  &  $C^p=1.5$ & $C^p=2.0$  \\
 $T_{\Lambda}$        
& 8.56(3)   & 8.97(3)   &  9.83(3) &10.46(4) 
& 8.23(3)   & 8.76(3)   &  9.38(3) &10.02(4)
& 7.94(3)   & 8.39(3)   &  9.01(3) & 9.58(3)  \\
 $v_0(r)(1-\varepsilon)$ 
& -16.08(6) & -16.36(6) & -16.78(6) & -17.03(6)  
& -13.69(5) & -14.07(5) & -14.37(5) & -14.51(5) 
& -11.55(4) & -11.91(4) & -12.06(4) & -12.22(5)  \\
$v_0(r)\varepsilon P_x$ 
& -1.56(1)  & -1.58(1)  & -1.61(1) &  -1.63(1)
& -2.97(1)  & -3.04(1)  & -3.09(1) &  -3.11(1)
& -4.27(2)  & -4.40(2)  & -4.43(2) &  -4.47(2)  \\
 $(\frac{1}{4})v_\sigma T_\pi^2(r) \mbox{\boldmath$\sigma$}_\Lambda\cdot\mbox{
\boldmath$\sigma$}_i$ 
&  0.014(0)  & 0.015(0)   &  0.017(0) & 0.017(0)   
&  0.012(0)  & 0.013(0)   &  0.014(0) & 0.014(0)
&  0.009(0)  & 0.010(0)   &  0.011(0) & 0.011(0)  \\
 $v_{\Lambda i}$  
& -17.63(6) &-17.93(6)  & -18.38(6)  & -18.65(7) 
& -16.65(6) &-17.10(6)  & -17.42(6)  & -17.61(7) 
& -15.80(6) &-16.30(6)  & -16.48(6)  & -16.68(7)  \\
 $V_{\Lambda ij} ^D$   
& 2.49(1) & 4.60(1)  &  7.99(4)  & 12.17(6)
& 2.16(1) & 4.20(1)  &  7.35(4)  & 11.53(6) 
& 1.80(1) & 3.81(1)  &  6.81(4)  & 10.65(5)  \\
 $V_{\Lambda ij}^{ P}$ 
& -1.36(1)  & -4.40(2) & -10.02(4)& -15.99(6) 
& -1.36(1)  & -4.54(2) & -9.42(4) & -15.35(6) 
& -1.28(1)  & -4.16(2) & -8.97(4) & -14.38(6)   \\
 $V_{\Lambda ij}^{ S}$   
&-0.025(0)  &  0.017(1) & 0.066(1) & 0.100(1)
&-0.025(0)  &  0.009(1) & 0.057(1) & 0.087(1) 
&-0.030(1)  &  0.009(1) & 0.050(1) & 0.077(1)   \\
 $V_{\Lambda ij}^{2\pi}$ $= V_{\Lambda ij}^{P} + V_{\Lambda ij}^{ S}$ 
& -1.38(1)  & -4.39(2) & -9.95(4) &  -15.89(6)
& -1.39(1)  & -4.52(2) & -9.36(4) &  -15.26(6)
& -1.31(1)  & -4.16(2) & -8.92(4) &  -14.30(6)   \\
$V_{\Lambda ij}=V_{\Lambda ij}^D +V_{\Lambda ij}^{2\pi}$ 
&  1.11(1)   & 0.21(2)    & -1.96(2) &  -3.73(3)
&  0.77(1)   & -0.33(2)   & -2.05(2) &  -3.73(3)
&  0.49(1)   & -0.34(2)   & -2.11(2) &  -3.65(3)   \\
$V_\Lambda{}=v_{\Lambda i}+V_{\Lambda ij}$ 
& -16.52(6)   &  -17.72(6)  & -20.34(6) & -22.38(7)
& -15.88(6)   &  -17.43(6)  & -19.43(6) & -21.34(7)
& -15.31(6)   &  -16.64(6)  & -18.59(6) & -20.33(7)   \\
$E_\Lambda=T_{\Lambda}+ V_\Lambda^{}$
 & -7.96(3)  &  -8.75(4)  &-10.51(4) &  -11.92(5)
 & -7.64(4)  &  -8.67(4)  &-10.05(4) &  -11.32(5)
 & -7.37(4)  &  -8.26(4)  & -9.58(4) &  -10.75(5)   \\
$T_{NC}$     
&  117.59(15)     &  118.42(15)  & 118.69(15) & 119.26(15)
&  116.97(15)     &  117.84(15)  & 118.42(15) & 118.84(15)
&  117.51(15)     &  118.43(15)  & 118.07(15) & 118.53(15)  \\ 
$v_{NN}$     
& -134.65(14)    & -134.53(14)  & -133.18(14) & -132.55(14)
& -134.36(14)    & -134.05(14)  & -133.41(14) & -132.79(14)
& -134.99(14)    & -134.85(14)  & -133.42(14) & -132.87(14)  \\
$V_{NNN}$    
&  -5.81(2)      & -5.98(2)   & -5.84(2) & -5.66(2)
&  -5.82(2)      & -5.97(2)   & -5.83(2) & -5.67(2)
&  -5.99(2)      & -6.15(2)   & -5.93(2) & -5.76(2) \\
 $V_{NC}=v_{ij}+V_{ijk}$  
& -140.46(14)  & -140.51(14)  & -139.02(14)  & -138.21(14) 
& -140.16(14)  & -140.03(14)  & -139.24(14)  & -138.46(14) 
& -140.98(14)  & -141.00(14)  & -139.35(14)  & -138.63(14)   \\
$E_{NC}=T_{NC}+V_{NC}$  
&  -22.88(4)      &  -22.10(5)  &-20.32(4) & -18.95(5)
&  -23.21(4)      &  -22.18(5)  &-20.82(4) & -19.52(5)
&  -23.47(4)      &  -22.58(5)  &-21.27(4) & -20.11(5) \\
$E=E_\Lambda +E_{NC}$ 
&-30.84(2) & -30.85(2)  &-30.84(3)  & -30.87(4)
&-30.85(2) & -30.85(2)  &-30.86(3)  & -30.84(4)
&-30.84(2) & -30.84(2)  &-30.86(3)  & -30.86(4) \\ 
\end{tabular}
\end{ruledtabular}
\end{table}
\end{turnpage}
\begin{turnpage}
\begin{table}
\caption{\label{tab2} Energy breakdown for the set 
$\overline{v}2$ ($\overline{v}=6.10$ and $v_\sigma=0.24$).  
All quantities are in units of MeV except for $\varepsilon$.
}
\begin{ruledtabular}
\begin{tabular}{lcccccccccccc}
   &\multicolumn{4}{c}{ $\varepsilon=0.1$ }
   &\multicolumn{4}{c}{ $\varepsilon=0.2$ }
   &\multicolumn{4}{c}{ $\varepsilon=0.3$ } \\
 &   $C^p=.5$   &  $C^p=1.0$  &  $C^p=1.5$ & $C^p=2.0$  
 &   $C^p=.5$   &  $C^p=1.0$  &  $C^p=1.5$ & $C^p=2.0$  
 &   $C^p=.5$   &  $C^p=1.0$  &  $C^p=1.5$ & $C^p=2.0$  \\
\hline
 $T_{\Lambda}$        
& 8.09(3)   & 8.54(3)   &  9.19(3) & 9.91(3)
& 7.80(3)   & 8.15(3)   &  8.65(3) & 9.51(3)
& 7.57(3)   & 7.94(3)   &  8.44(3) & 9.18(3)  \\
 $v_0(r)(1-\varepsilon)$ 
& -14.61(5) & -14.79(5) & -15.13(5) & -15.48(5)
& -12.54(5) & -12.64(5) & -12.94(5) & -13.30(5) 
& -10.59(4) & -10.73(4) & -11.04(5) & -11.15(4)  \\
$v_0(r)\varepsilon P_x$ 
& -1.41(1)  & -1.42(1)  & -1.45(1)  &  -1.47(1)
& -2.70(1)  & -2.72(1)  & -2.77(1) &  -2.83(1)
& -3.88(2)  & -3.93(2)  & -4.03(2) &  -4.03(2)  \\
 $(\frac{1}{4})v_\sigma T_\pi^2(r) \mbox{\boldmath$\sigma$}_\Lambda\cdot\mbox{
\boldmath$\sigma$}_i$ 
&  0.007(0)  & 0.007(0)   &  0.008(0) & 0.009(0)  
&  0.007(0)  & 0.008(0)   &  0.008(0) & 0.009(0)
&  0.005(0)  & 0.005(0)   &  0.006(0) & 0.007(0)  \\
 $v_{\Lambda i}$  
& -16.02(6) &-16.20(6)  & -16.57(6) & -16.94(6)
& -15.23(6) &-15.35(6)  & -15.70(6) & -16.12(6) 
& -14.46(6) &-14.65(6)  & -15.04(6) & -15.18(6)  \\
 $V_{\Lambda ij} ^D$   
& 1.65(1) & 3.46(2)  &  6.43(4)  &  10.39(5)
& 1.32(1) & 3.09(2)  &  5.95(4)  &   9.81(5) 
& 0.99(1) & 2.71(2)  &  5.47(4)  &   9.22(5)  \\
 $V_{\Lambda ij}^{ P}$ 
& -1.33(1)  & -4.36(2) & -9.00(4) &  -14.95(6)
& -1.23(1)  & -4.04(2) & -8.08(3)  &  -14.26(6) 
& -1.19(1)  & -3.98(2) & -7.61(3)  &  -13.84(6)   \\
 $V_{\Lambda ij}^{ S}$   
&-0.028(0)  &  0.014(1) & 0.049(1) & 0.087(0)
&-0.035(0)  &  0.007(0) & 0.040(1) & 0.078(0) 
&-0.032(0)  &  0.009(0) & 0.036(1) & 0.074(0)   \\
 $V_{\Lambda ij}^{2\pi}$ $= V_{\Lambda ij}^{P} + V_{\Lambda ij}^{ S}$ 
& -1.35(1)  & -4.35(2) & -8.59(4) &  -14.86(6)
& -1.27(1)  & -4.04(2) & -8.04(4) &  -14.18(6)
& -1.22(1)  & -3.97(2) & -7.80(4) &  -13.77(6)   \\
$V_{\Lambda ij}=V_{\Lambda ij}^D +V_{\Lambda ij}^{2\pi}$ 
& -0.30(1)   & -0.88(2)   & -2.53(2) &  -4.47(3)
&  0.05(1)   & -0.94(2)   & -2.09(2) &  -4.37(3)
&  0.49(1)   & -1.26(2)   & -2.33(2) &  -4.55(3)   \\
$V_\Lambda{}=v_{\Lambda i}+V_{\Lambda ij}$ 
& -15.72(6)   &  -17.08(6)  & -19.10(6) & -21.41(6)
& -15.18(6)   &  -16.29(6)  & -17.79(6) & -20.49(6)
& -14.69(6)   &  -15.91(6)  & -17.37(6) & -19.73(6)   \\
$E_\Lambda=T_{\Lambda}+ V_\Lambda^{}$
 & -7.62(3)  &  -8.54(4)  & -9.90(4) & -11.51(5)
 & -7.37(3)  &  -8.14(4)  & -9.15(4) & -10.98(5)
 & -7.13(4)  &  -7.97(4)  & -8.93(4) & -10.54(5)   \\
$T_{NC}$     
&  116.82(15)     &  116.53(15)  & 117.36(15) & 118.09(15)
&  116.73(15)     &  116.68(15)  & 117.19(15) & 117.55(15)
&  116.81(15)     &  117.03(15)  & 117.16(15) & 117.40(15)  \\ 
$v_{NN}$     
& -134.22(14)    & -132.90(14)  & -132.59(14) & -131.80(14)
& -134.40(14)    & -133.50(14)  & -133.12(14) & -131.82(14)
& -134.43(14)    & -133.72(14)  & -133.12(14) & -131.96(14)  \\
$V_{NNN}$    
&  -5.83(2)      & -5.92(2)   & -5.74(2) & -5.62(2)
&  -5.81(2)      & -5.90(2)   & -5.78(2) & -5.60(2)
&  -6.10(2)      & -6.21(2)   & -5.94(2) & -5.73(2) \\
 $V_{NC}=v_{ij}+V_{ijk}$  
& -140.05(14)  & -138.82(15)  & -139.31(15)  & -137.42(14)
& -140.21(14)  & -139.41(14)  & -138.90(15)  & -137.42(14) 
& -140.53(14)  & -139.93(15)  & -133.12(15)  & -137.69(14)   \\
$E_{NC}=T_{NC}+V_{NC}$  
&  -23.22(4)      &  -22.31(5)  &-20.97(5)  & -19.32(5)
&  -23.47(4)      &  -22.71(5)  &-21.71(5) & -19.87(5)
&  -23.71(4)      &  -22.89(4)  &-21.91(4) & -20.29(5) \\
$E=E_\Lambda +E_{NC}$ 
&-30.84(2) & -30.85(2)  &-30.87(2)  & -30.84(3)
&-30.85(2) & -30.85(2)  &-30.86(2)  & -30.85(3)
&-30.84(2) & -30.87(2)  &-30.84(2)  & -30.84(3) \\ 
\end{tabular}
\end{ruledtabular}
\end{table}
\end{turnpage}
\begin{turnpage}
\begin{table}
\caption{\label{tab3} Energy breakdown for the set 
$\overline{v}3$ ($\overline{v}=6.05$ and $v_\sigma=0.24$).  
All quantities are in units of MeV except for $\varepsilon$.
}
\begin{ruledtabular}
\begin{tabular}{lcccccccccccc}
   &\multicolumn{4}{c}{ $\varepsilon=0.1$ }
   &\multicolumn{4}{c}{ $\varepsilon=0.2$ }
   &\multicolumn{4}{c}{ $\varepsilon=0.3$ } \\
 &   $C^p=.5$   &  $C^p=1.0$  &  $C^p=1.5$ & $C^p=2.0$  
 &   $C^p=.5$   &  $C^p=1.0$  &  $C^p=1.5$ & $C^p=2.0$  
 &   $C^p=.5$   &  $C^p=1.0$  &  $C^p=1.5$ & $C^p=2.0$  \\
\hline
 $T_{\Lambda}$        
& 7.60(3)   & 8.09(3)   &  8.66(3) & 9.18(3)
& 7.27(3)   & 7.73(3)   &  8.14(3) & 8.87(3)
& 7.11(3)   & 7.45(3)   &  7.90(3) & 8.48(3)  \\
 $v_0(r)(1-\varepsilon)$ 
& -13.26(5) & -13.57(5) & -13.69(5) & -13.95(5)
& -11.20(4) & -11.48(5) & -11.64(5) & -12.02(5) 
& -9.58(4)  & -9.68(4)  &  -9.90(4) & -9.94(4)  \\
$v_0(r)\varepsilon P_x$ 
& -1.27(1)  & -1.30(1)  & -1.30(1) &  -1.31(1)
& -2.40(1)  & -2.45(1)  & -2.48(1) &  -2.54(1)
& -3.49(2)  & -3.51(2)  & -3.56(2) &  -3.58(2)  \\
 $(\frac{1}{4})v_\sigma T_\pi^2(r) \mbox{\boldmath$\sigma$}_\Lambda\cdot\mbox{
\boldmath$\sigma$}_i$ 
& -0.003(0)  &-0.002(0)   & -0.002(0) & -0.001(0)
&  0.003(0)  & 0.004(0)   &  0.005(0) &  0.006(0)
&  0.002(0)  & 0.002(0)   &  0.003(0) &  0.003(0)  \\
 $v_{\Lambda i}$  
& -14.53(6) &-14.87(6)  & -15.00(6)  & -15.27(6)
& -13.60(6) &-13.92(6)  & -14.10(6)  & -14.56(6) 
& -13.06(6) &-13.19(6)  & -13.48(6)  & -13.52(6)  \\
 $V_{\Lambda ij} ^D$   
& 0.87(0) & 2.61(1)  &  5.30(3)  &  8.79(5)
& 0.51(0) & 2.16(1)  &  4.68(3)  &  8.25(5) 
& 0.24(0) & 1.82(1)  &  4.31(3)  &  7.50(4)  \\
 $V_{\Lambda ij}^{ P}$ 
& -1.20(1)  & -4.07(2) & -8.53(4)  &  -13.45(6)
& -1.11(1)  & -3.84(2) & -7.49(4)  &  -12.87(6) 
& -1.06(1)  & -3.62(2) & -7.24(4)  &  -12.27(6)   \\
 $V_{\Lambda ij}^{ S}$   
&-0.032(0)  &  0.010(0) & 0.046(0) & 0.071(1)
&-0.037(0)  &  0.003(0) & 0.030(0) & 0.062(1) 
&-0.037(0)  &  0.001(0) & 0.029(0) & 0.060(1)   \\
 $V_{\Lambda ij}^{2\pi}$ $= V_{\Lambda ij}^{P} + V_{\Lambda ij}^{ S}$ 
& -1.23(1)  & -4.06(2) & -6.48(4) &  -13.37(6)
& -1.15(1)  & -3.84(2) & -7.46(4) &  -12.81(6)
& -1.09(1)  & -3.63(2) & -7.21(4) &  -12.21(6)   \\
$V_{\Lambda ij}=V_{\Lambda ij}^D +V_{\Lambda ij}^{2\pi}$ 
& -0.37(1)   & -1.44(2)   & -3.19(2) &  -4.59(3)
& -0.64(1)   & -1.68(2)   & -2.78(2) &  -4.56(3)
& -0.85(1)   & -1.81(2)   & -2.90(2) &  -4.71(3)   \\
$V_\Lambda{}=v_{\Lambda i}+V_{\Lambda ij}$ 
& -14.89(7)   &  -16.32(6)  & -18.19(6) & -19.86(6)
& -14.24(7)   &  -15.60(6)  & -16.89(6) & -19.12(6)
& -13.91(7)   &  -11.48(6)  & -16.38(6) & -18.23(6)   \\
$E_\Lambda=T_{\Lambda}+ V_\Lambda^{}$
 & -7.29(3)  &  -8.23(3)  & -9.52(4) & -10.67(4)
 & -6.97(3)  &  -7.87(4)  & -8.75(4) & -10.25(4)
 & -6.80(3)  &  -7.55(4)  & -8.48(4) &  -9.74(4)   \\
$T_{NC}$     
&  115.97(15)     &  116.33(15)  & 116.54(15) & 116.42(15)
&  115.28(15)     &  115.70(15)  & 116.01(15) & 116.81(15)
&  116.04(15)     &  116.28(15)  & 116.23(15) & 115.68(15)  \\ 
$v_{NN}$     
& -133.81(14)    & -133.07(14)  & -132.06(14) & -131.98(14)
& -133.44(14)    & -132.81(14)  & -132.32(14) & -131.73(14)
& -134.04(14)    & -133.41(14)  & -132.63(14) & -131.08(14)  \\
$V_{NNN}$    
&  -5.73(2)      & -5.88(2)   & -5.80(2) & -5.61(2)
&  -5.73(2)      & -5.87(2)   & -5.75(2) & -5.67(2)
&  -6.03(2)      & -6.16(2)   & -5.95(2) & -5.71(2) \\
 $V_{NC}=v_{ij}+V_{ijk}$  
& -139.53(14)  & -138.95(14)  & -137.86(14)  & -136.59(14)
& -139.17(14)  & -138.68(14)  & -138.07(14)  & -137.40(14) 
& -140.08(14)  & -139.57(14)  & -138.58(14)  & -136.79(14)   \\
$E_{NC}=T_{NC}+V_{NC}$  
&  -23.57(4)      &  -22.61(5)  &-21.32(4)  &-20.17(5)
&  -23.89(4)      &  -22.98(4)  &-22.09(4) & -20.59(5)
&  -24.04(4)      &  -23.29(4)  &-22.36(4) & -21.11(5) \\
$E=E_\Lambda +E_{NC}$ 
&-30.85(2) & -30.84(2)  &-30.84(2)  & -30.84(3)
&-30.85(2) & -30.85(2)  &-30.84(2)  & -30.84(3)
&-30.84(2) & -30.84(2)  &-30.85(2)  & -30.86(3) \\ 
\end{tabular}
\end{ruledtabular}
\end{table}
\end{turnpage}
\begin{table} 
\caption{\label{tab5}
Variation of the slope, $\partial W^D/\partial \varepsilon$, 
with $C^P$ and  $\overline{v}$.
}
\begin{ruledtabular}
\begin{tabular}{lccc}
 $C^P$ (MeV) &  $\overline{v}1$ (MeV) &   $\overline{v}2$ (MeV) & $\overline{v}3$ (MeV) \\
\hline
0.5 &  -0.016(1) & -0.017(1)  & -0.017(1)    \\
1.0 &  -0.017(1) & -0.019(1)  & -0.019(1)     \\
1.5 &  -0.019(1) & -0.022(1)  & -0.023(1)    \\
2.0 &  -0.021(1) & -0.023(1)  & -0.024(1)     \\
2.5 &  -0.022(1) & -0.025(1)  & -0.026(1)     \\
\end{tabular}
\end{ruledtabular}
\end{table}


\begin{thebibliography} {299}
\bibitem{AAU06} A. A. Usmani, Phys. Rev. {\bf C73}, 011302(R) (2006).
\bibitem{AAU206} A. A. Usmani and Z. Hasan, submitted to Phys. Rev. {\bf C}.
\bibitem{AV18} R. B. Wiringa, V. G. J. Stoks, and R. Schiavilla, Phys.
              Rev. {\bf C51}, 38 (1995).
\bibitem{NNN-IX} B. S. Pudliner, V. R. Pandharipande, J. Carlson
                and R. B. Wiringa, Phys. Rev. Lett. {\bf 74}, 4396 (1995).
\bibitem{NNN} J. Carlson,  V. R. Pandharipande and R. B. Wiringa, 
               Nucl. Phys.  {\bf A401}, 59 (1983).
\bibitem{DHT72} R. H. Dalitz, R. C. Herndon and Y. C. Tang, Nucl. Phys.
                {\bf B47}, 109 (1972).
\bibitem{Gal75} A. Gal, Adv. Nucl. Phys. {\bf 8}, 1 (1975).
\bibitem{Hungerford84} E. V. Hungerford and L. C. Biedenhorn, Phys. Lett. 
                   {\bf 142B}, 232 (1984).
\bibitem{Ahn01} K. Ahn {\it et al.}, Phys. Rev. Lett {\bf 87}, 
                 132504 (2001).
\bibitem{Tyapkin} A. A. Tyapkin, Sov. J. Nucl. Phys. {\bf 22}, 89 (1976).
\bibitem{Dover} C. B. Dover and S. H. Kahana, Phys. Rev. Lett. {\bf 39}, 1506 
                (1977).
\bibitem{Khanna} K. Tsushima and F. C. Khanna, Phys. Rev. {\bf C67}, 015211 
                  (2003).
\bibitem{BR70} A. R. Bodmer, D. M. Rote and A. L. Mazza, 
                  Phys. Rev. {\bf C2}, 1623 (1970). 
\bibitem{BR71} A. R. Bodmer and D. M. Rote
                  Nucl. Phys. {\bf A169}, 1 (1971).
\bibitem{Rozynek79} J. Rozynek and J. Dabrowski, 
                  Phys. Rev. {\bf C20}, 1612 (1979); {\bf 23}, 1706(1981);
                  Y. Yamamoto and H. Bando, Suppl. Prog. Theor. Phys. Suppl. 
                  {\bf 81}, 9(1985); Y. Yamamoto, Nucl. Phys. {\bf A450}, 275c 
                  (1986). 
\bibitem{UB99} Q. N. Usmani and A. R. Bodmer,
                 Phys. Rev. {\bf C60}, 055215  (1999).
\bibitem{BU88} A. R. Bodmer and Q. N. Usmani,
                 Nucl. Phys. {\bf A477}, 621 (1988).
\bibitem{Bhaduri67} R. K. Bhaduri, B. A. Loiseau and Y. Nogami,
                 Anns. Phys. (N. Y) {\bf 44}, 57 (1967).
\bibitem{Friar99} J. L. Friar, D. H\"{u}ber and U. van Kolck, 
                Phys. Rev. {\bf C59}, 53 (1999).
\bibitem{Coon79} S. A. Coon, M. D. Scadron, P. C. McNamee, B. R. Barrett,
                 D. W. E. Blatt and B. H. J. McKellar, Nucl. Phys. {\bf A317}, 
                 242 (1979).
\bibitem{Coelho83} H. T. Coelho, T. K. Das and M. R. Robilotta,
                 Phys. Rev. {\bf C28}, 1812 (1983); M. R. Robilotta and
                 H. T. Coelho, Nucl. Phys. {\bf A460}, 645 (1986).
\bibitem{Eden93} J. A. Eden and M. F. Gari, Phys. Rev. {\bf C53}, 1510 (1996).
\bibitem{Kolck94}C. Ord\'{o}\~{n}ez and U. van Kolck, 
                Phys. Lett. B {\bf 291}, 459 (1992); U. van Kolck, Phys. Rev.
                {\bf C49}, 2932 (1994).
\bibitem{Fujita57} J.-I Fujita and H. Miyazawa, Prog. Theor. Phys. {\bf 17}, 
                360 (1957).
\bibitem{SCP01} S. C. Pieper, V. R. Pandharipande, R. B. wiringa and 
                J. Carlson, Phys.  Rev. {\bf C64}, 014001 (2001).
\bibitem{BUC84} A. R. Bodmer and Q. N. Usmani and J. Carlson, 
                  Phys. Rev. {\bf C29}, 684 (1984).
\bibitem{Lagaris81} I. E. Lagaris and V. R. Pandharipande,
                 Nucl. Phys. {\bf A359}, 331 (1981).
\bibitem{AAU03} A. A. Usmani and S. Murtaza, Phys. Rev. {\bf C68}, 
                024001 (2003).
\bibitem{UPU95} A. A. Usmani, S. C. Pieper and Q. N. Usmani,
                 Phys. Rev. {\bf C51}, 2347 (1995).
\bibitem{AAU95} A. A. Usmani, Phys. Rev. {\bf C52}, 1773, (1995).
\end{thebibliography}
\end{document}